\documentclass{aastex61}
\received{\today}
\revised{---}
\accepted{---}
\submitjournal{ApJ}

\shorttitle{ Kinematics of an isolated galaxy, CIG 993}
\shortauthors{C\'ardenas-Mart\'inez \& Fuentes-Carrera}

\begin{document}

\title{Kinematics of the isolated luminous infrared galaxy CIG 993}

\correspondingauthor{Nelli C\'ardenas-Mart\'inez}
\email{ncardenas@esfm.ipn.mx, nelli\_cardenas@hotmail.com}

\author{Nelli C\'ardenas-Mart\'inez}
\affiliation{Escuela  Superior  de  F\'isica  y  Matem\'aticas,  Instituto  Polit\'ecnico  Nacional  (ESFM-IPN),\\
U.P.  Adolfo  L\'opez  Mateos,  edifico  9, Zacatenco, CP 07730 Mexico City, Mexico}

\author{Isaura Fuentes-Carrera}
\affiliation{Escuela  Superior  de  F\'isica  y  Matem\'aticas,  Instituto  Polit\'ecnico  Nacional  (ESFM-IPN),\\
U.P.  Adolfo  L\'opez  Mateos,  edifico  9, Zacatenco, CP 07730 Mexico City, Mexico}




\begin{abstract}

We present scanning Fabry-Perot interferometric observations of CIG 993, an apparently isolated luminous infrared galaxy also exhibiting luminous blue compact and Wolf-Rayet galaxy features, as well as high star formation rate.
Our high resolution observations of the H$\alpha$ emission line allowed us to derive the radial velocity field, as well as  the velocity dispersion $\sigma$, and the residual velocity fields of the galaxy. This galaxy exhibits   several kinematical components.
On one hand, the velocity gradients detected on the velocity field can be associated with a rotating disk -contrary to previous results with less spectral resolution-.  However, the velocity field, the $\sigma$ and residual velocity field show significant deviations from circular motions in the central part of the galaxy that matches a region with high number of Wolf-Rayet and O stars, coincident with the blue luminous component of the galaxy. We find narrow and broad velocity components for the ionized gas in the central part of the galaxy. The  broad component is evidence of a central outflow related with the on-going burst of stellar formation. 
The morpho-kinematical analysis of the galaxy indicates we are only seeing the brightest parts of the galaxy which correspond to the bulge, a central bar and the beginning of the disk.
We believe CIG 993 is a disk galaxy harboring important star-forming processes most likely caused by a relatively recent interaction. 
This could imply that small encounters could change the global characteristics of a galaxy without disturbing the main rotation disk motion, nor the  morphology of the galaxy.

\end{abstract}




	\section{Introduction} \label{sec:intro}

\defcitealias{rc31991}{RC3}

Luminous Infrared Galaxies (LIRGs) are galaxies whose  far infrared (FIR) emission is larger than the emission at all the other wavelengths combined, FIR luminosity lies between $\rm 10^{11}L_{\bigodot}$ and $\rm 10^{12}L_{\bigodot}$ \citep{S&M1996}. Although it seems clear that their FIR luminosities excess  arises from the thermal heating of dust, the nature of the engine behind this heating has two main candidates: AGN activity, or an extreme star forming event, or a combination of both phenomena \citep{Veilleux1995,Yuan2010,Alonso-Herrero2012}.

In the  Local Universe, 
\cite{Ishida2004} found that  $100\%$  of LIRGs with $\rm L_{IR}>10^{11.5}L_{\bigodot}$ shows at least some evidence for tidal features and that $70\%$ of LIRGs with $\rm10^{11.1}L_{\bigodot}<L_{IR}<10^{11.5}L_{\bigodot}$  also tend to be disturbed. 

Several works have shown that in the  Local Universe,  a high star formation rate (SFR) is primarily triggered by interactions or mergers \citep{Kennicutt1987,Brassington2015}. It can be said that there is a relation between current or past interactions and the LIRG phenomena.  However, detailed multiwavelength  studies of local LIRGs already identified that not all LIRGs show obvious evidence of interactions \citep{S&M1996,Hann2011}. These exceptions to the merger hypothesis  may  represent the end-stage of the merger after most of the obvious tidal features have disappeared, or an alternative way of producing LIGs, 
such as  minor  mergers  and  bar  instabilities that will  tend to  drive  star  formation \citep{Melbourne2008,Hann2011}.

 In order to identify the processes behind these galaxies we study the kinematics  of one of them: an apparently isolated LIRG that has not experienced any  interaction or merger and without detected nuclear activity. Isolated galaxy samples should provide the  baseline for interpreting the high SFR in LIRGs that are not mergers, due to the fact that isolated galaxies are minimally affected by the environment (neighbor galaxies). In these sense, perhaps the best compilation of isolated galaxies available is the Catalog of Isolated Galaxies \citep[CIG,][]{Kara73} and the Analysis of the interstellar Medium of Isolated GAlaxies \citep[AMIGA,][]{AMIGA05}. 
The \cite{Kara73} Catalog was original assembled with the requirement that no similar sized galaxies with diameter D (between 1/4 and 4 times diameter D of the CIG galaxy) lie within a radius of 20D, but some CIG galaxies have neighbors that were later identified by \cite{Verley2007} and \cite{Argurdo-Fernandez2013}.

In  this  paper we focus on the optical kinematic properties  of CIG 993.  In  Section 2, a brief bibliographical review on this  galaxy  is presented. The observations and data reduction are presented in Section 3. In Section 4, we present the derived velocity field, velocity dispersion field, emission profile decomposition, rotation curve of the galaxy, and the residual velocity field, paying special attention to the role of non-circular motions. In Section 5, we present a morphological analysis of the galaxy at different wavelengths.  In Section 6, we discuss the link  between galaxy properties at different wavelengths and the kinematic features. Finally, the conclusions are shown in Section 7.  In  this paper,  we  adopt $H_0 = 71$  km s$^{-1}$ Mpc$^{-1}$, $\Omega_M=0.27$,  and $\Omega_\Lambda=0.73$.

\section{CIG 993}
\label{sec:CIG 993}

\begin{figure}
\epsscale{1.2}
\plotone{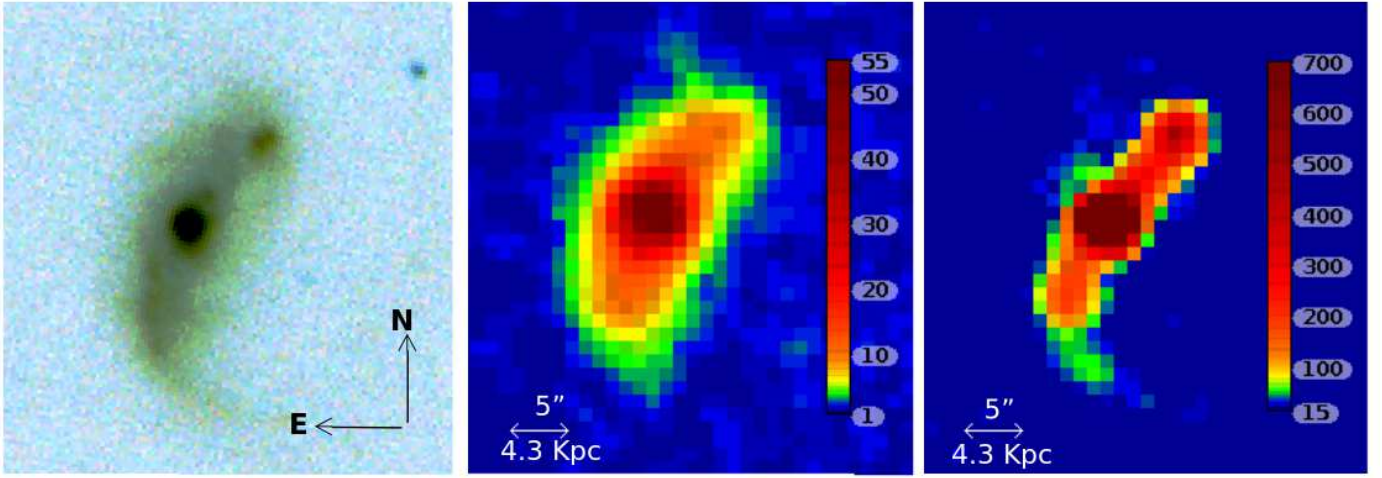}
\caption{From left to right:  SDSS composite image  of CIG 993 in the g'r'i'-bands,  continuum  image and monochromatic $H\alpha$  image from PUMA observations. 
All the images in this paper have the same orientation and correspond to a box of $41^{\prime\prime} \times44^{\prime\prime}$. The color scale units are counts. \label{fig:mono-cont}}
\end{figure}

Left panel of  Fig.\ref{fig:mono-cont} shows CIG 993, also called Mrk 309,  a spiral galaxy classified as Sa  \citep[][hereafter RC3]{rc31991} and as Sc \citep{Sulentic2006}, with an heliocentic velocity of 12636 km s$^{-1} $  (\citetalias{rc31991}). General parameters of the galaxy are shown in Table \ref{tab:Prop}. Some other properties of the galaxy  are summarized in the following subsections.

\begin{deluxetable*}{lll}[b!]
\tablecaption{General Parameters of CIG 993  \label{tab:Prop}}
\tablecolumns{3}
\tablenum{1}
\tablewidth{0pt}
\tablehead{
\colhead{Parameter} &
\colhead{Value} &
\colhead{Reference} 
}
\startdata
Galaxy name			&CIG 993, Mrk 309	& NED\tablenotemark{a} \\
RA (J2000)			&$\rm 22^{h}52^{m}34.7^{s}$ 		&NED\tablenotemark{a}\\
DEC (J200)			&$\rm +24 ^{\circ} 43^{\prime} 50^{\prime\prime}$			&NED\tablenotemark{a}\\
Hubble Type			& Sa				&\citetalias{rc31991}\\
					& Sc				&\cite{Sulentic2006}\\
Nuclear activity	& HII				&\cite{PoggiantiWu2000}\\
Redshift			&  0.042149			& \citetalias{rc31991}\\
Velocity			&  12636 km s$^{-1}$	&\citetalias{rc31991}\\	
Distance			& 183.9 $\pm 0.1$ Mpc		&\cite{Fernandez2010}\\
Major axis\tablenotemark{b}&  $32.9 ^{\prime \prime }$	& \citetalias{rc31991}\\ 
Minor axis\tablenotemark{b}&  $16.9 ^{\prime \prime }$	& \citetalias{rc31991}\\
$\rm L_{FIR}$		& $10^{11.72} \rm L_{\bigodot}$	&\cite{PoggiantiWu2000}\\
					& $10^{11.24} \rm L_{\bigodot}$	&\cite{Lisenfeld2007} \\
					& $10^{11.50} \rm L_{\bigodot}$	&\cite{Fernandez2010}\\					
$\rm SFR_{H\alpha}$	&	6.7 $\rm M_{\bigodot}yr^{-1}$		&\cite{PoggiantiWu2000}\\
$\rm SFR_{FIR}$		&	90.9 $\rm M_{\bigodot}yr^{-1}$		& \cite{PoggiantiWu2000}\\
					&	54.52 $\rm M_{\bigodot}yr^{-1}$	& This work\\
\enddata
\tablenotetext{a}{The NASA/IPAC Extragalactic Database (NED) is operated by the Jet  Propulsion  Laboratory,  California  Institute  of  Technology,  under contract with the National Aeronautics and Space Administration.}
\tablenotetext{b}{Measured up to 25.0 mag isophote in B-band}
\end{deluxetable*}

\subsection{LIRG}	
CIG 993  is a LIRG because of its high FIR luminosity. In this work we follow  the  convention  established in \cite{S&M1996}, in which the FIR luminosity ($\rm L_{FIR}$)  corresponds to a spectral range between 40 - 500 $\mu m$.  \cite{ Lisenfeld2007}  and \cite{Fernandez2010} reported $L_{FIR}$ for CIG 993 of $\rm 10^{11.24} L_{\bigodot}$ and $\rm 10^{11.5}L_{\bigodot}$, respectively. Both groups  used the \cite{S&M1996} equation to determine $\rm L_{FIR}$  in the range between 40 - 500 $\mu m$, but the values differ because of the redshift determination distance (i.e. adopted $H_0$ value) and the method to derive flux density from the IRAS observations, \cite{ Lisenfeld2007} derive flux estimation  from the best fitting point source temperature and the maximum flux density within the signal range, while \cite{Fernandez2010} took the flux density from the IRAS Point Source Catalog. 

\cite{PoggiantiWu2000} also derived the FIR luminosity but in the range of 42.2 to 122.5 $\mu m$ using the \cite{Dennefeld1986} equation finding a FIR luminosity of $\rm 10^{11.72}L_{\bigodot}$.

\subsection{Isolation}
Although, CIG 993 is an isolated galaxy \citep{Kara73},  20 neighbor galaxies are reported in the  vicinity of the galaxy according to the neighbors catalog of the AMIGA  project \citep{Verley2007}. Only one neighbor lies within a radius of twenty times the diameter  of CIG 993 having a similar size. This could be a background galaxy or a truly interacting neighbor, but because of the lack of redshift measurements this fact  is unknown.


\subsection{Nuclear Activity}

In a spectroscopy study of galaxies with ultraviolet (UV)  continuum, \cite{Afanasev1980} classified CIG 993 as a  Seyfert 2 galaxy. \cite{Sabater2008} used the  FIR color criterion \citep{deGrijp1985} in which galaxies hosting an AGN have in general a flatter spectrum in  FIR,  to propose CIG 993 as an AGN candidate. However in the same article, \cite{Sabater2008} use the q-parameter \citep{Helou1985} that relates the radio and the FIR emission, and according to this method, the galaxy is not an AGN. \cite{OHalloran2005} also did IR diagnostics with ISO fluxes. Considering that the ratio of the 7.7 $\mu m$ flux to the continuum level at this wavelength  can provide a measure of the level of activity within the nucleus \citep{Genzel1998}, they concluded that CIG 993 hosts a compact burst of star formation and that an AGN is not required within the central burst \citep{OHalloran2005}.

Other optical diagnostic tests 
by \cite{OsterbrockDahari1983} and \cite{PoggiantiWu2000} concluded that CIG 993 is clearly an object in which the nuclear gas is photoionized by O stars. Following the previous results we assume that the galaxy does not host an AGN.

\subsection{e(c) class}

\cite{PoggiantiWu2000}  did  optical spectroscopy  of CIG 993 and  classified it as an e(c) \citep{Dressler1999}. The e(c) spectrum was identified because of its weak to moderate Balmer lines in absorption, an equivalent width (EW) of [O II] $3727{\rm\AA}$ in emission shorter than $40{\rm\AA}$ and $H\delta$ emission  shorter than $4{\rm\AA}$. 

The e(c) spectra are similar to those  in present-day spirals \citep{Poggianti1999}. However there is an alternative way of reproducing an e(c) spectrum: starbursts of sufficiently long duration ($\gg$0.1 Gyr) can in their later phases mimic the e(c) classes. After an initial e(b) phase, in which the spectrum displays  very strong emission lines associated with a galaxy that is undergoing strong star formation \citep{Dressler1999}, the starbursting galaxy displays an e(c) spectrum as long as the 
star formation continues \citep{Poggianti1999}.


\subsection{Luminous compact blue galaxies (LCBGs)}

According to  \cite{Pisano2001} and \cite{Garland2004} criterion, CIG 993 is a LCBG because its absolute blue magnitude ($M_B$) is brighter than −18.5 mag in a effective radii of 4$ ^{\prime \prime }$, its effective surface brightness is brighter than 21 mag arcsec$^{-2}$ in the B band and  its rest-frame $B $ - $V$ colour is bluer than 0.6 mag \citep{Perez-Gallego2011}.
  
The LCBG class is mostly populated by a morphological mixture of starburst galaxies with a compact and highly ionized central zone \citep{Garland2004}. However  \cite{MallenOrnelas1999} sample of LCBG   can be identified with bright irregulars or late-type spirals.

\subsection{Wolf-Rayet galaxy}

A striking feature in the spectrum of CIG 993 is the presence of a broad emission band at a rest wavelength of $\lambda \approx 4680 \,{\rm\AA}$ \citep{Osterbrock&Cohen1982, Kunth&1986, Fernandes2004}.  They interpret this band as arising from the contribution in the integrated spectra of about $10^{4}$ Wolf-Rayet (WR) stars  whose emission seem to come largely if not entirely from the center of the galaxy \citep{Osterbrock&Cohen1982, Fernandes2004}. 
By using $H\beta$ emission, \cite{Osterbrock&Cohen1982}, \cite{Kunth&1986} and \cite{Fernandes2004} concluded that CIG 993 hosts the same number of O stars than WR stars implying that a massive star formation episode must have occurred recently, about the time required for an O star to evolve to a WR star, roughly $10^7$ yr ago and over a time fairly short in comparison with this age \citep{Osterbrock&Cohen1982}.  This point will be discussed in Section \ref{sec:displaymath2}.


\subsection{Kinematics - Previous works}

\cite{Perez-Gallego2011} used three-dimensional optical spectroscopy observations to study the kinematics of CIG 993. Objects from the \cite{Perez-Gallego2011} sample were observed using the PMAS spectrograph \citep{Roth2005} in the PPAK mode \citep{Kelz2006} at the Calar Alto 3.5-m telescope.  The spectrum was centered at $5316{\rm\AA}$ and included $H\alpha$ emission of the galaxy. The configuration  used provided a nominal spectral resolution of $10.7{\rm\AA}$  FWHM ($\sigma \sim 255$ km s$^{-1}$ at $5316{\rm\AA}$). \cite{Perez-Gallego2011} derived the velocity field of this galaxy using different dithering positions and  spatially binned the data (coadding fibres to achieve a minimum signal-to-noise of 13) in the outer area of the galaxy.  CIG 993 was classified as a galaxy with complex kinematics, meaning that neither the velocity map nor the velocity width map are compatible with regular disk rotation. Those authors also found that the velocity map is misaligned  by $\sim90^\circ$ with the morphological major axis  \citep[Figure 4 in][]{Perez-Gallego2011}.

Additionally, \cite{MirabelSanders1988} and \cite{Fernandez2010} studied the HI profile of CIG 993 at a resolution of 8.3 km s$^{-1}$ and 23.8 km s$^{-1}$, respectively. Both used the Arecibo Observatory 305m radio telescope  and found that CIG 993 profiles are asymmetric. In their works, these authors suggested that asymmetric profiles can be due to dynamical disruptions of the disks of spiral galaxies, due to interactions. However, \cite{Espada2011} found a lack of a strong correlation between $\rm L_{FIR}$ and asymmetric HI flux in the AMIGA sample, 
 i.e.,  that  induced star formation caused by possible interactions in the AMIGA sample is small compared  to  that  from  secular  evolution.  Still,  there  is  an  excess of about 10\% of asymmetric profiles for the LIRGs in that sample. This might be linked to recent accretion events in a small number of CIG galaxies \citep{Espada2011}.

\section{Observations and Data Reduction} \label{sec:floats}

Observations of CIG 993 were done on November 2014 at the f/7.5 Cassegrain focus of the  2.1 m telescope at the Observatorio Astron\'omico Nacional of San Pedro M\'artir (OAN-SPM), Mexico, using the scanning Fabry-Perot interferometer PUMA \citep{Rosado1995}. We used a $2048\times2048$ Marconi 2 CCD detector, with a binning factor of four, resulting in a field of view of $512 \times 512$ pixels with a spatial sampling of $1.27 ^{\prime \prime }$/pixel. 

 A  filter, centered in $\lambda_0 = 6819 \,{\rm\AA}$ and  with a bandwidth of $86 \,{\rm\AA} $,  was selected according to the recession velocity of the galaxy. We obtained one object interferogram  with 48 channels  and two calibration interferograms, producing velocity cubes of $512\times 512 \times 48$, with 180 s exposure time per channel, covering a full spectral range of $21.54\,{\rm\AA}$ or the equivalent of 945 km s$^{-1}$ with a finesse of $\sim 24$.  
  
In order to calibrate the $H\alpha$ cube we used a neon lamp whose $ 6717.04 \,{\rm\AA} $ line was the   closest to the redshifted nebular wavelength to scan the same 48 channels. These were later used for phase and wavelength calibration of the data cubes.  Observational  and instrumental parameters are listed in Table \ref{tab:Obs}.

The data were processed using the  ADHOCw\footnote{The manual is available at \textit{www.astro.umontreal.ca/fantomm/Modedemploi/ADHOC\_manuel.pdf}} (developed by J. Boulesteix) and IRAF\footnote{IRAF is distributed by National Optical Astronomy Observatory, operated by the Association of Universities for Research in Astronomic, Inc., under cooperative agreement with the National Science Foundation.} softwares. Standard corrections were done in each cube (like  removal of cosmic rays, subtraction  of  bias, etc.).

 The calibration in wavelength was fixed for each profile per pixel using the calibration cube. After the calibration of the raw data cube for which the surfaces of constant wavelength are paraboloids, the cube was converted to a Cartesian cuboid of 48 planes, each $512 \times 512$ pixels separated by $0.45 {\rm\AA}$. Once the wavelength calibration was done, the OH sky lines were subtracted using as reference Fig. 13 from  \cite{Osterbrock1996}

\setcounter{table}{1}
\begin{table}[h!]
\renewcommand{\thetable}{\arabic{table}}
\centering
\caption{Observational and Instrumental Parameters  } \label{tab:Obs}
\begin{tabular}{ll}
\tablewidth{0pt}
\hline
\hline
Parameter & Value \\
\hline
\decimals
Instrument				&  PUMA					\\

Size detector			& $2048\times2048 \, pixels$\\
Image scale ($4\times4$ binning) & $1.27 ^{\prime \prime }/pixel$\\
Detector					& Marconi 2				\\ 
\hline
Finesse  ($\mathcal F$)			    		&  $\sim 24$    			\\
Order of interference 	&  317  for $H\alpha$   				\\
Free spectral range ($\mathcal FSR$)		&  $21.54\,{\rm\AA}$   (945.5 km s$^{-1}$)  	\\
Spectral Resolution  	&  $0.45\,{\rm\AA}   $  (19.7 km s$^{-1}$)  	\\
FP Scanning Steps		& 48						\\

\hline
\multicolumn{2}{c}{ }
\end{tabular}
\end{table}

Once this process was done, we obtained for each pixel an intensity value (counts) for each channel. The intensity profile found along the scanning process contains information about the monochromatic $H\alpha$ emission and continuum emission of CIG 993. The continuum was obtained by considering the mean of the 3 lowest intensities of the 48 channels. The monochromatic value for each panel was obtained by integrating the monochromatic profile. Both maps are shown in Fig. \ref{fig:mono-cont}. 

\begin{figure}
\plotone{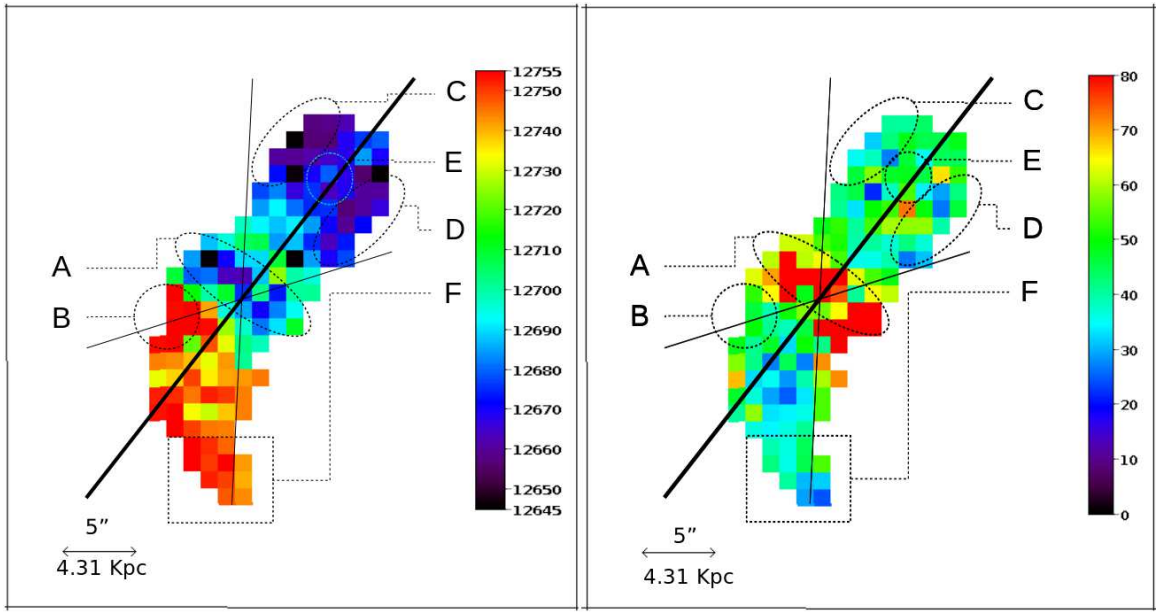}
\caption{ $Left$: Observed variable resolution velocity field (VF) for CIG 993. $Right$: Velocity dispersion field for CIG 993. The color scale units are in km s$^{-1}$. The thick line represents the kinematic major axis and the thin line represent the sectors used to derive the rotation curve (RC) in Section \ref{sec:RC}. 
\label{fig:VIT-sigma}}
\end{figure}

Finally, the velocity maps  were derived by computing the barycenter of the $H\alpha$ profile for each pixel. We performed a spectral gaussian smoothing ($\sigma= 2.54^{\prime \prime }$) and in order to get a sufficient signal-to-noise ratio on the outer parts of the galaxy  we performed  three spatial gaussian smoothings ($\sigma=0.9, \, 1.35, \, 1.8 {\rm\AA}$), so that a variable resolution radial velocity map was built using less spatially and spectrally smoothed pixels for regions with an originally higher signal-to-noise ratio  and more spatially and  spectrally smoothed pixels for the outer parts of the galaxy (see left panel of Fig. \ref{fig:VIT-sigma})

Positional astrometry was performed by comparing the positions of the brightest stars in the Fabry-Perot field with the positions in the SDSS images and USNO catalog yielding maximum uncertainties in position of 0.5$^{\prime \prime }$.

\section{Kinematic Results}

\subsection{Velocity Field}
\label{sec:VF}

Left panel of  Fig. \ref{fig:VIT-sigma} shows the velocity field (VF) of CIG 993.
The largest velocity values are seen on south-southeastern side of the galaxy; the lowest velocity values are seen on the extreme of northern side, along with a region in the central parts of the galaxy that seems to lie perpendicular to the main axis of the galaxy indicated with region A in Fig.\ref{fig:VIT-sigma}.
Globally, a velocity gradient is seen from low velocity values on the NW side of the galaxy to the SE side where large velocity values are seen.
This type of gradient is associated with a rotating disk, the so-called ''spider diagram''.
However, important contributions of both low and high velocities are seen in the central parts of the galaxy (regions A and B in left panel of  Fig. \ref{fig:VIT-sigma}).
These will be discussed in the following sections.
Also, in the NW side of the galaxy. a particular pattern of velocities is seen close to the NW tip of the galaxy.
Instead of the usual velocity distribution associated with a rotating disk, the VF of CIG 993 shows a region (region E in  left panel of  Fig. \ref{fig:VIT-sigma}) with larger velocities that the regions on one side and the other of the main axis of the galaxy (regions C and D).
The velocity distributions seen in regions C and D correspond more to the spider diagram velocity distribution than the velocities in region E.
The velocities along the apparent warp of the galaxy (region F) show no particular distribution other than the one that is expected on a spider diagram.

\subsection{Velocity Dispersion Field}\label{sec:FWHM}

The velocity dispersion, $\sigma$, for each pixel was computed from the H$\alpha$ velocity profiles after correction from instrumental and thermal widths,
$\sigma_{inst} = FSR / (\mathcal F \times 2\sqrt {2 \ln 2} ) =$ 16.4 \
km \ s$^{-1}$ and $\sigma_{th} =$ 9.1 \  km \ s$^{-1}$, respectively. The latter value was estimated by assuming an electronic temperature of T$_{e}$ = 10$^4$ K  \citep{Spitzer1978,Osterbrock1989} in the expression $\sigma_{th}$=(kT$_{e}$/m$_{H}$)$^{1/2}$.
Assuming that all the profiles (instrumental, thermal, and turbulent) are described by gaussians, the velocity dispersion was estimated using $ \sigma = (\sigma_{obs}^2 - \sigma_{inst}^2 - \sigma_{th}^2)^{1/2}$.
Right panel of Fig.\ref{fig:VIT-sigma} shows the velocity dispersion field of CIG 993.
The largest velocity dispersion values (160 km s$^{-1}$) are seen for pixels in the central region and its surroundings, particularly along the minor axis of the galaxy (region A in Fig.\ref{fig:VIT-sigma}). The velocity dispersion values in the northern side of the galaxy are in average 20 km s$^{-1}$ larger than those in the southern half, especially in region E. 

\subsection{Emission Profiles}
\label{Sec:EmissionProf}

\begin{figure}
\epsscale{1.0}
\plotone{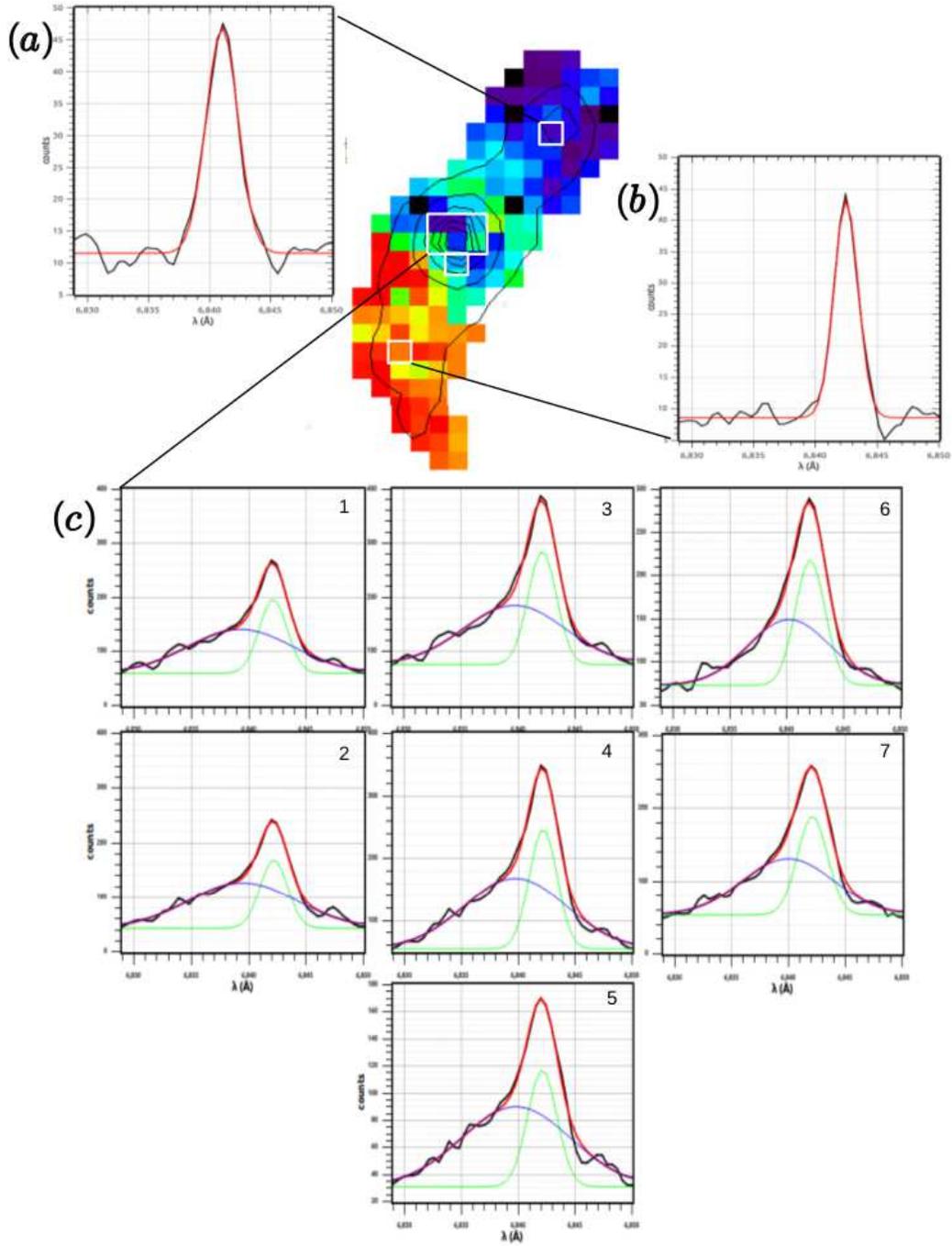} 
\caption{  H$\alpha$ emission profiles in CIG\,993 obtained with PUMA, taken from the north \textit{(a)},  south \textit{(b)} and central \textit{(c)} parts of the galaxy. Original profiles (black line) were fitted with narrow component (green). In the central parts of the galaxy, a broad component  (blue) was also considered for the fit.
  For these profiles, the best fit is shown in red lines.
  The VF of CIG 993 is shown in the central upper part of the figure. Isophotes correspond to the SDSS images in the u'-band \label{fig:profiles}}
\end{figure}

Figure \ref{fig:profiles} shows the emission profiles from different regions in CIG 993; all of them show the $H\alpha$  emission line. 
Inspection of the emission profiles of CIG 993 showed the presence of composite profiles in the central parts of the galaxy (panel (c) in Figure \ref{fig:profiles}).
The location of these profiles matches the region on the velocity dispersion map with the largest values (region A in Fig.\ref{fig:VIT-sigma}).
The composite profiles in the central parts of the galaxy were fitted with a narrow and a broad gaussian components. Values for each component are given in Table \ref{tab:components}.
No composite profiles are seen on the northern or southern sides of the galaxy (panels (a) and (b) in Fig.\ref{fig:VIT-sigma}).

\startlongtable
\begin{deluxetable}{cllll}
\tablecaption{Parameters value for the narrow and broad $H\alpha$ components.
\label{tab:components}}
\tablehead{
\colhead{Pixel} & \colhead{Preak} & \colhead{Center} & \colhead{Width}&    \colhead{Height}\\
\colhead{} & \colhead{} & \colhead{[km s$^{-1}$]}& \colhead{[km s$^{-1}$]}& \colhead{[counts]}
}
\startdata
1    &    1& 12733& 105& 136 \\
(284,239) &    2& 12603& 402& 80    \\
\hline
2    &    1& 12735& 104& 125    \\
(284,240) &    2& 12612& 413& 83    \\
\hline            
3    &1& 12733& 108& 208        \\
(285,239)  &2& 12622& 389& 109        \\
\hline        
4    &1& 12735& 107& 192        \\
(285,240) &2& 12624& 393& 113        \\
\hline        
5    &1& 12732& 112& 86        \\
(285,241)    &2&    12627& 421& 58        \\
\hline        
6    &1& 12730& 114& 144        \\
(286,239)  &2& 12646& 315& 75        \\
\hline        
7    &1& 12732& 115& 136        \\
(286,240)   &2& 12636& 352& 77        \\
\enddata

\end{deluxetable}

\subsection{Rotation Curve}
\label{sec:RC}

\begin{figure}
\plotone{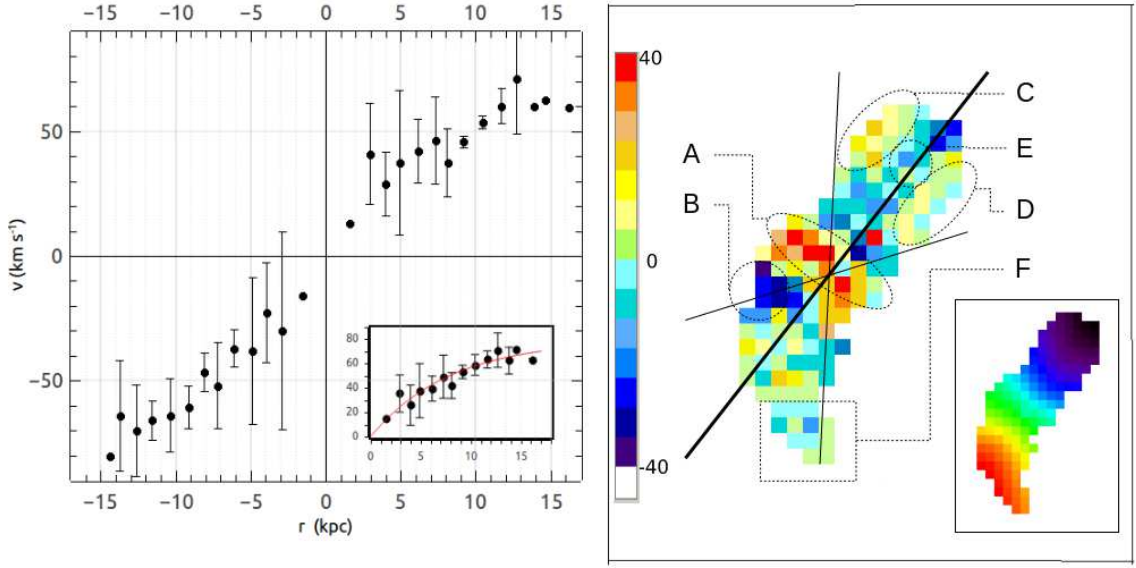}
\caption{  $Left$: Rotation curve for CIG 993 chosen in order to obtain a symmetric flat curve and to minimize the scatter on each side of the curve  for the outer parts of the galaxy, ``H$\alpha$ RC'' in Section \ref{sec:RC}. Best analytical fit is shown with a solid line in the inset figure. $Right$: Residual velocities for CIG\,993 after subtraction of a rotational model obtained from the observed gaseous velocity field. $Inset$: Modeled velocity field for CIG\,993, the color scale is the same than in the  velocity field of Fig. \ref{fig:VIT-sigma} . The color scale units are in km s$^{-1}$. The thick line represents the kinematic major axis and the thin line represent the sectors used to derive the RC.
\label{fig:cr_Halpha}}
\end{figure}

\begin{figure}
\plotone{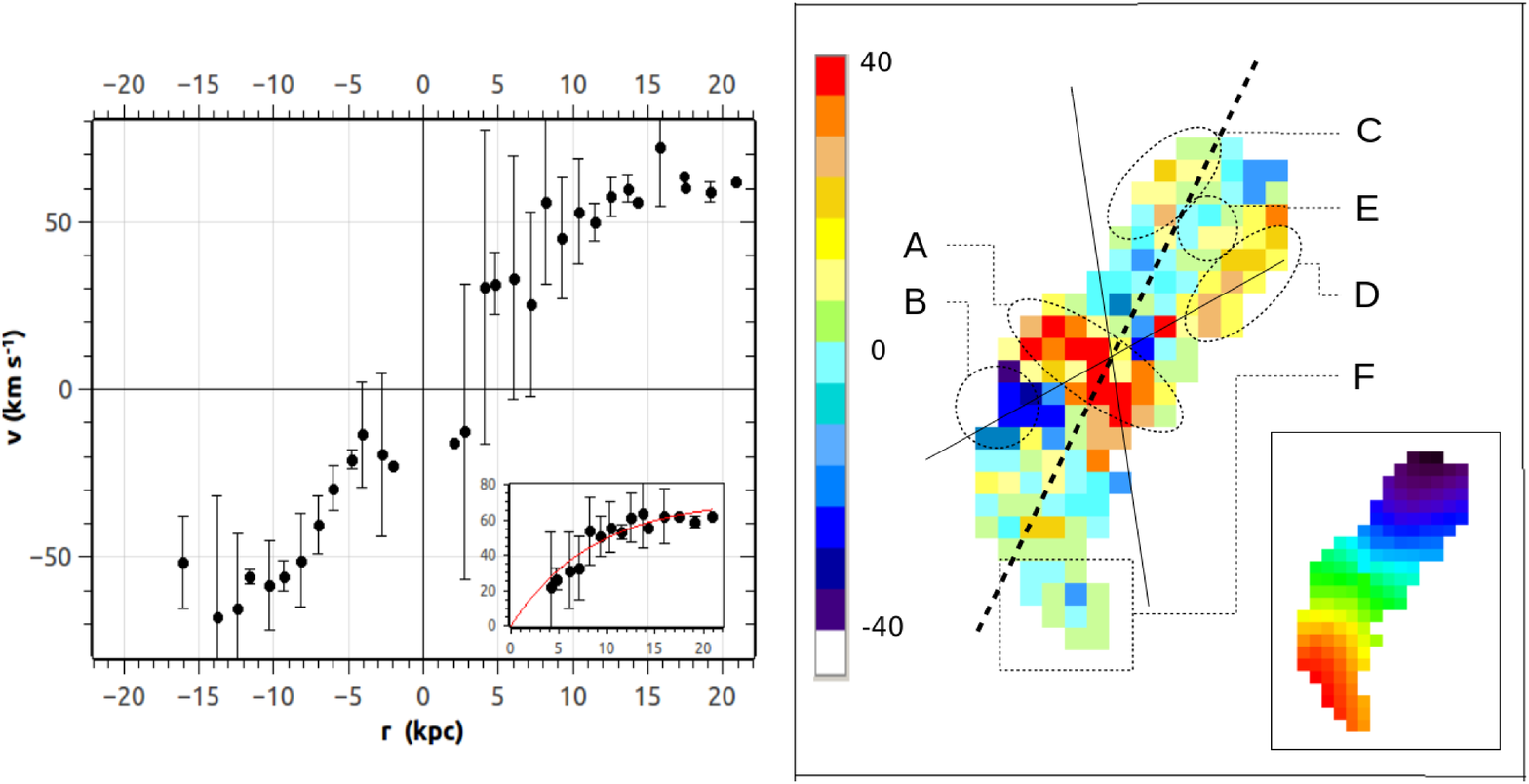}
\caption{   $Left$: Rotation curve for CIG 993 obtained using parameters derived from the K-band image. The ``K-band RC'' in Section \ref{sec:RC}. Best analytical fit is shown with a solid line in the inset figure. $Right$: Residual velocities for CIG\,993 after subtraction of a rotational model obtained from the observed gaseous velocity field. $Inset$: Modeled velocity field for CIG\,993, the color scale is the same than in the  velocity field of Fig. \ref{fig:VIT-sigma} . The color scale units are in km s$^{-1}$. The thick line represents the kinematic major axis and the thin line represent the sectors used to derive the RC.
\label{fig:cr_Kband}}
\end{figure}

In Section \ref{sec:VF}, we pointed out the different velocity components of CIG 993 that do not seem to follow the rotation motion of a disk (regions A, B and E in Fig.\ref{fig:VIT-sigma}). Still, a global velocity gradient can be seen from one extreme of the galaxy to the other.
We thus derived a  rotation curve (RC)  from the kinematics of the ionized gas of the galaxy.
The RC of the galaxy is derived from the  VF, which gives the projected velocity along the line of sight in each position of the galaxy. The RC of the galaxy was obtained with  the ADHOCw software. In order to plot the RC, we obtained a velocity value at a radius $r$ by considering an elliptical ring of one pixel  width ($1.27^{\prime\prime}$) whose main axis equals $r$ and whose inclination matches the inclination of the galaxy disk.  
Considering the VF of CIG 993 is perturbed in different regions, we derived two RCs taking into account different considerations:

\begin{enumerate}

	\item  Kinematic parameters were chosen  in order to obtain a symmetric flat curve  for all points on the VF inside a certain angular  sector, and to minimize the scatter on each side of the curve. The following set of values was used:  kinematical center  (RA=$\rm 22^{h} 52^{m} 34.72^{s}$, DEC=$+24^{\circ} 43^{\prime}48.5^{\prime\prime}$) -which matches the photometric $H\alpha$ center within $1^{\prime\prime}$ (0.86 kpc)-,  inclination angle of the galaxy $ i=58.5^\circ \pm1^\circ$ and position angle of the major axis  (PA=+322$\pm5^\circ$, measured from the north to the east). 	Additional parameters include the systemic velocity of the galaxy $V_{sys}=12715$ km s$^{-1}$ and an angular sector of $35^\circ$ on each side of the major axis  (see Fig \ref{fig:VIT-sigma}). Only points within these sectors are considered for the derivation of the RC in order to reduce the dispersion.  Uncertainties in the parameters were computed by varying each parameter and evaluating the symmetry and dispersion of the RC.
The derived RC is shown in Fig.\ref{fig:cr_Halpha}. Table \ref{tab:RC} summarizes these results. The maximum rotation velocity derived from these values is 80$\pm 8$ km s$^{-1}$.
We shall call this RC, the ``H$\alpha$ RC''.

	\item Considering that the central parts of the galaxy are undergoing a strong starburst and that the profile decomposition in this region has shown the presence of large velocity dispersions and of two velocity components (Figs. \ref{fig:VIT-sigma} and \ref{fig:profiles}), it might not be a reliable assumption to seek to symmetrize the inner parts of the galaxy in order to derive the RC.
For this reason, we derived a second RC based on the parameters given by the K-band image (see bottom middle panel of Fig. \ref{fig:imagenes}) which traces the old stellar disk of the galaxy. The RC was then derived considering as kinematical center the photometric center of the K-band image (RA=$\rm 22^{h} 52^{m} 34.72^{s}$, DEC=$+24^{\circ} 43^{\prime} 50^{\prime\prime}$), and the inclination and P.A. were derived from the fit of an ellipse to the outermost isophotes of this image ($ i=68.0^\circ $ and PA=+333.5$^\circ$, respectively). The photometric center of the K band (see Fig. \ref{fig:imagenes}) is shifted 1 pix in the PUMA images (1.27 arcsec = 1.09 kpc) from the center used to derive the ``H$\alpha$ RC''. Fig.\ref{fig:cr_Kband} shows the RC derived with these assumptions. Table \ref{tab:RC} shows  these results.
We shall call this RC,  the ``K-band RC''.

\end{enumerate}

The inserts plotted in Fig. \ref{fig:cr_Halpha} and \ref{fig:cr_Kband} show the global RC that was derived by averaging the velocities on both sides of the galaxy (approaching and receding sides) for a single radius  in each case. The red line is the best fit using  minimum chi-square estimation and the parametric function $V(r)=A[1-\exp{(-r/b)}]$ where A and b are constants.  Their values are presented in Table \ref{tab:RC}.

The ``H$\alpha$ RC'' is less dispersed on both sides of the galaxy than the ``K-band RC''. Large error bars ($\sim$ 46 km s$^{-1}$) are seen for the ``K-band RC'' from 3 to 7 kpc for the receding side of the galaxy.
The same occurs for the average RC, the average ``K-band RC'' shows larger error bars than the ``H$\alpha$ RC'' up to 15 kpc.
The fit of the average RCs show no significant difference for the $A$ value within the error bars.
The turning point of the fitted RC, $b$, is slightly shorter for the ``K-band RC''.


\begin{deluxetable*}{lcc}[b!]
\tablecaption{Kinematical Parameters \label{tab:RC}}
\tablecolumns{7}
\tablewidth{0pt}
\tablehead{
\colhead{Parameter} &\colhead{$H\alpha$ RC} &\colhead{K-band RC}
}
\startdata
Systemic  Velocity ($V_{sys}$)    &12715 km s$^{-1}$                &12709 km s$^{-1}$                \\
Inclination     ($i$)            &+58.5$\pm1^\circ$                &+68.8$\pm8^\circ$                \\
Position angle (PA)                &+322$\pm5^\circ$                &+333.5$\pm3^\circ$                \\
Center (RA,DEC)                    &($\rm 22^{h} 52^{m} 34.72^{s}, +24^{\circ} 43^{\prime}48.5^{\prime\prime}$)         &($\rm 22^{h} 52^{m} 34.72^{s}$, $+24^{\circ} 43^{\prime} 50^{\prime\prime}$)\\
Angular sector                    & $35^\circ$                    &$35^\circ$                        \\
A\tablenotemark{a}                & $80\pm8$    km s$^{-1}$            &$71\pm6$    km s$^{-1}$            \\
$1/b$\tablenotemark{a}            & $0.138\pm0.02$                &$0.117\pm0.02$                    \\
\enddata
\tablenotetext{a}{The best fit using the parametric function $V(r)=A[1-\exp{(-r/b)}]$}
\end{deluxetable*}

\subsection{Model velocity map and residual velocity filed} 

The presence of non-circular motions within the galactic disk was explored by deriving the residual velocities map, which is obtained by subtracting a rotation model velocity field from the observed velocity field. This model is a two-dimensional projection of the analytical fit of the observed RC assuming there are only uniform circular motions in the plane of the galaxy \citep{Iqui2004}.  These fields prove to be very useful to evaluate the validity of the parameters chosen to compute the rotation curve of a disk galaxy \citep{Warner1973}.

Using each RC, we derived the residual velocities field for CIG 993.
These are shown in the right panels of Figures \ref{fig:cr_Halpha} and \ref{fig:cr_Kband}.
In none of the residual velocity fields do we find a residual velocity pattern that could be related to an error in the determination of the kinematics parameters \citep{Warner1973}, so that the different velocities displayed in the galaxy correspond to actual non-circular motions of the gas.
Both residual maps (right panel of Fig.\ref{fig:cr_Halpha} and \ref{fig:cr_Kband}) show net residual velocities close to zero,  mainly outside the 4 kpc central radius. Considering the spectral resolution of our observations (see Table \ref{tab:Obs}), residual velocity values of $\pm$10 km s$^{-1}$ can be considered close to zero, implying the predominance of circular motions in $\sim 60\%$ of the galaxy especially in the outer parts where ionized gas is detected.
Important non-circular motions are seen in both residual velocity fields along the minor axis of the galaxy (region A). 
In the central region (inner 4 kpc), there  are important velocity differences, that go from  -37  to +43 km s$^{-1}$  for the H$\alpha$ residual velocity field, and from -38  to +75 km s$^{-1}$ for the K-band residual velocity field
These residual velocities could be associated with gas ejected from the active star formation in the central parts of the galaxy \citep{Veilleux2005,OparinMoiseev2015}.
For both maps, residual velocity values close to zero are seen in the SW tip of the galaxy (region E in Figures \ref{fig:cr_Halpha} and \ref{fig:cr_Kband}).

Though both residual velocity fields are quite similar,   
the residual velocity distribution seems less symmetrical for the K-based residual velocity field, especially regarding regions C and D, which have similar velocity values in both the VF and the dispersion velocity map (Fig. \ref{fig:VIT-sigma}).
The RC and residual velocity field  that best describe the rotational motion of CIG 993 will be discussed in Section \ref{sec:displaymath2}.

\section{CIG\,993 in different wavelengths}
\label{sec:displaymath}

\begin{figure}
\epsscale{1.15}
\plotone{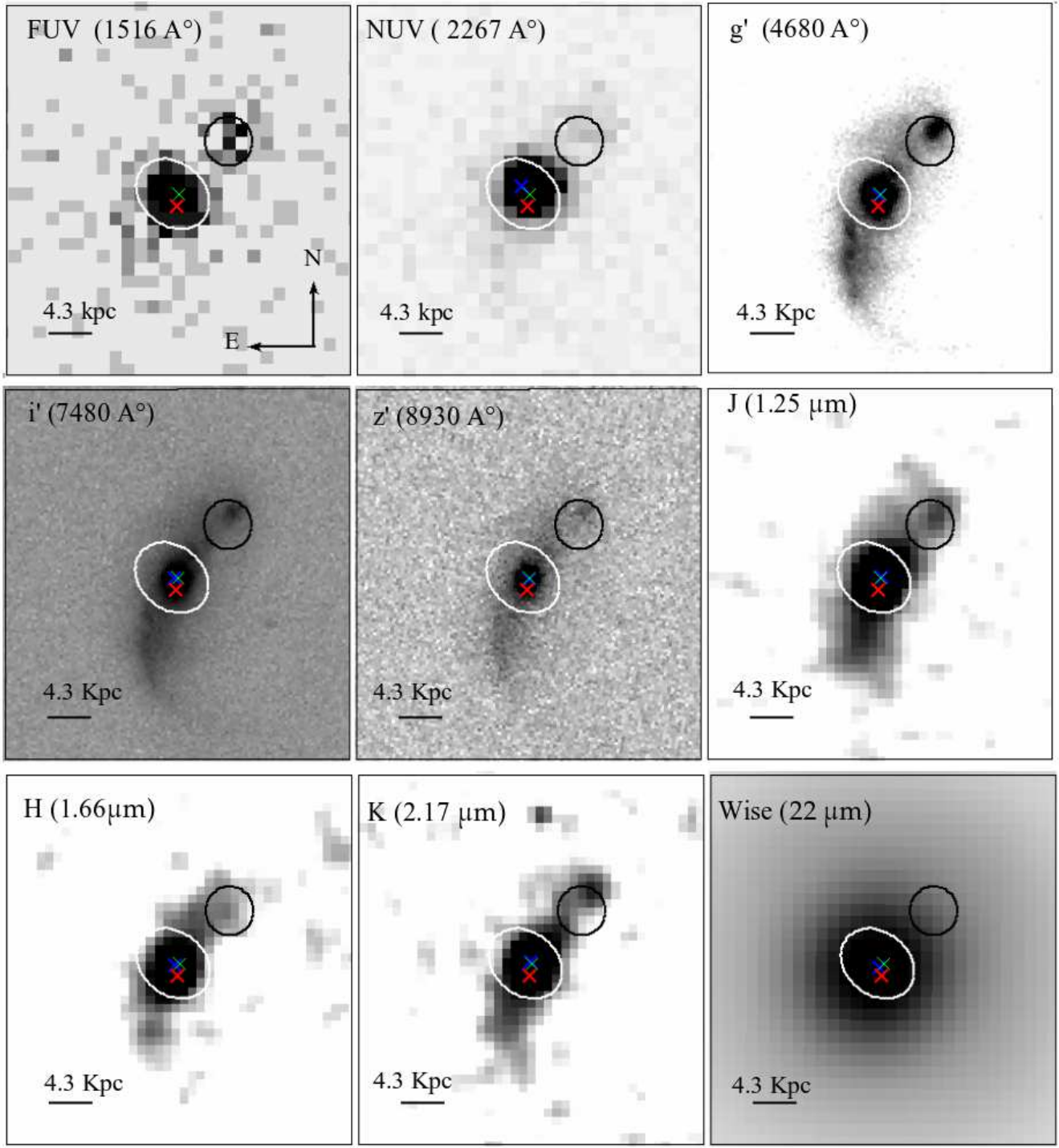} 
\caption{Images for CIG\,993 in different wavelengths. \textit{From left to right and top to bottom}: FUV images from GALEX, NUV images from GALEX, B-band from DSS, i' and z'-band SDSS images, J, H and K-band from 2MASS and WISE image at 22 $\mu m$. Images have the same scale  and orientation in a box of $41^{\prime\prime}\times 44^{\prime\prime}$.  The green cross on each image represents the photometric center of the FUV image, blue cosses represent the photometric center on each band and the red cross is the location of kinematical center  used to derive the H$\alpha$ RC -see Section \ref{sec:RC}.  \label{fig:imagenes}}
\end{figure}

Fig. \ref{fig:imagenes} shows images of CIG\,993 in different wavelengths in order to identify particular morphological features. The far and near ultraviolet (FUV and NIR) images from GALEX show bright emission in the central part of the galaxy. Emission is also seen in the northwestern  part of the galaxy (see black circle in Fig. \ref{fig:imagenes}). This feature, which is also visible in the $H \alpha$ images (see Fig. \ref{fig:mono-cont}), becomes less conspicuous as we move to redder wavelengths.

A central round-like ($\sim$ 3 kpc in radius) bright region is seen in all bands (see white ellipse in Fig. \ref{fig:imagenes}).
This feature could be associated with the bulge of the galaxy.
It is also detected in the UV bands because of the central starburst and the LBCG nature of CIG\,993.
Fig.\ref{fig:imagenes} shows that the location of the photometric center remains almost constant from band to band.  The green crosses in the images  represent the photometric FUV center and blue crosses indicate  the location of photometric center at each wavelength. 

A broad elongated arc-like feature is seen on both sides of the bulge, especially on the DSS B-band image, the SDSS images, and the 2MASS J-band image.
If this feature corresponds to the disk of CIG993, the southern side is more curved than the northern one, ressembling a warp.

ISO images at 7.7 and 14.3 $\mu m$ are reported  in Fig.1 of \cite{OHalloran2005}. The center is detected in both images.  At low flux levels the 7.7 $\mu m$ contours outline the disk of the galaxy, while the 14.3 $\mu m$ is more compact, just like the WISE images.

\section{Discussion} \label{sec:displaymath2}

The VF of CIG 993 shows several kinematical components indicated with letters A to F in Fig.\ref{fig:VIT-sigma}, \ref{fig:cr_Halpha} and \ref{fig:cr_Kband}.
These indicate the presence of non-circular motions in the central parts of the galaxy (regions A and B) and in the northwestern tip (region E), as well as circular motions implying disk-like rotation in regions C, D and F of the same Figure.
This implies that CIG 993 does harbor ``complex kinematics'' as \citet{Perez-Gallego2011} mentioned in their work, ``embedded'' in a field of rotational velocities which the spectral resolution of PUMA allows us to identify.
 Despite the \cite{Perez-Gallego2011} results, we find that the velocity map is aligned with the morphological major axis.

Making use of this spectral resolution, we fitted the H$\alpha$ emission profiles from the central region of the galaxy with a narrow and a broad components (see Fig. \ref{fig:profiles}). The mean  FWHM of the narrow  and broad components  are $\sim$109 km s$^{-1}$ and $\sim$384  km s$^{-1}$, respectively, which correspond to  $\sigma \sim$ 42  km s$^{-1}$ and $\sigma \sim$ 161  km s$^{-1}$.
The value of the narrow component is similar in average to the values found on the rest of the profiles on the galaxy (Fig.\ref{fig:VIT-sigma}).
The broad component is most likely associated with an outflow from the strong SF in the central parts of the galaxy (inner 4 kpc). 
The presence of ionized gas outflows in LIRGs seems to be common based on the detection of an $H\alpha$ broad line \citep{Arribas2014}.
The FWHM values of CIG\,993 seem to be in the range of the values that \cite{Arribas2014} find for the broad components in that type of outflows.
Those authors also point out that a central starburst is frequently associated with the presence of such ionized gas outflows.
As previously mentioned, CIG 993 is classified as an LCBG and shows WR features in its central parts.
In the same central region (see regions A and B in Fig. \ref{fig:VIT-sigma}), there is an important velocity difference in the residual velocity field; the difference in residual velocities goes from  -37  to +43 km s$^{-1}$, covering  $20\%$ of the observed galaxy (where the outflow is presumed to be).
This reinforces the fact that the kinematics in the central parts of the galaxy are highly disturbed from those of a rotating disk as a result of the  stellar winds coming from the central star-forming region. 
The ionized gas in the central parts of the galaxy is due to the combined influence of photoionization of young hot stars and  shocks produced by stellar winds due to the on-going star formation.

Two RCs were derived for CIG 993 under different assumptions: the first RC (``H$\alpha$ RC'') seeked to obtain the most symmetric and less dispersed RC from the kinematical information given by the VF; the second one (``K-band RC'') was computed using paramaters (center, $i$ and $P.A.$) derived from the K-band image -which best traces the old stellar disk of the galaxy.
The latter RC was derived in order to limit the effects of non-circular motions that the gas could have in the central parts of the RC due to the outflow detected in the VF, the $\sigma$ map and the emission profile decomposition.
A comparison of the RCs and of the associated residual velocity fields (Fig. \ref{fig:cr_Halpha} and \ref{fig:cr_Kband}) shows no major differences.
However the ``H$\alpha$ RC'' has less dispersion on both the approaching and receding sides than the ``K-band RC''.
Also, the resulting residual velocity field of the former is more symmetric and best represents the peculiar motions detected in both the VF and $\sigma$ map (regions C, D, and E in Fig.\ref{fig:VIT-sigma}).
For this reason, we consider the ``H$\alpha$ RC'' as the RC that best describes the rotation motion of the galaxy; even though non-circular motions are important in the central parts of the galaxy.
It is important to note that the regions associated with the outflow (regions A and B in  Fig.\ref{fig:VIT-sigma}) fall mostly outside the sector considered to compute the RC in both cases.
The adopted RC  does not reach  the flat part of the curve, but we estimated a maximum rotation velocity $V_{max}$ of $\sim$ 80 km s$^{-1}$ by using the limit in the RC fit.
The length of the curve is different for the approaching and receding sides, around 1 kpc larger for the approaching side.
The points in region F in Fig.\ref{fig:VIT-sigma} associated with the apparent warp follow the global behaviour of the RC and do not seem to be associated with any particular motion other than the rotation of the galaxy.
%

\begin{figure}
\epsscale{1.15}
\plotone{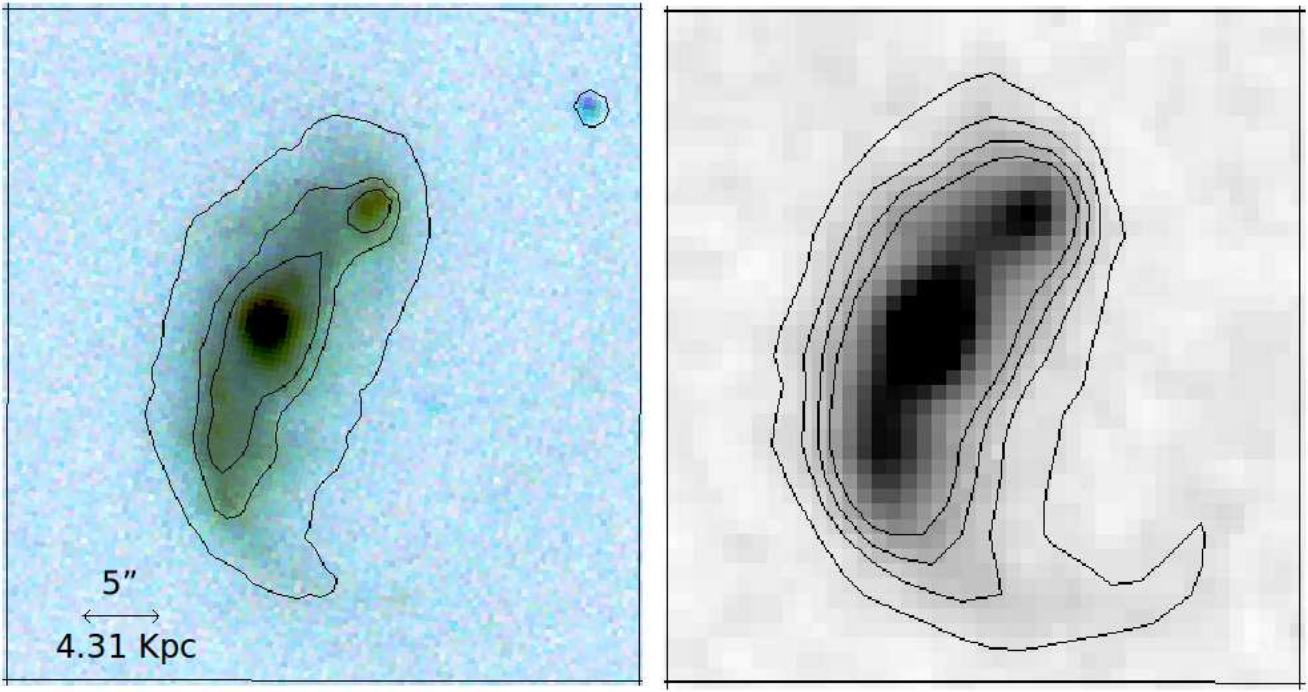} 
\caption{ {\it Left:} Composite  SDSS image of CIG 993 in the g'r'i'-band. {\it Right:}CIG 993 in  the DSS B-band. Isophotes corresponding to the faintest emission in each case.
  \label{fig:disk-features}}
\end{figure}

CIG\,993 is a LIRG galaxy that does not harbor nuclear activity which means that all the luminosity is due to the star formation phenomena. Using the conversion between FIR luminosity ($L_{FIR}=10^{11.5}L_{\bigodot}$) and  star formation rate (SFR)  given  in (Kennicutt 1998), we find that CIG\,993 has a SFR of 54.52 $M_{\bigodot}yr^{-1}$. This points to the starburst linked to the LCBG nature of the galaxy and the high number of WR and O  stars in the galaxy. 
Although a LIRG event needs a high fraction of dust covering the star formation and one might think that the two properties, LIRG and LCBG, cannot coexist, it is not odd to find systems with high FIR and blue luminosity \citep{Melbourne2005,Leitherer2013, Arribas2014} or  WR stars and FIR luminosity properties \citep{Armus1988}.  This seemingly paradoxical result is the consequence of their inhomogeneous ISM  as  \cite{Leitherer2013} suggest for LIRGs with important emission in UV band.
CIG 993 shows most of the emission in the B-band in the central parts of the galaxy (inner 4 kpc).
This is confirmed by the ultraviolet (GALEX), and $H\alpha$ (this work) images (left and central top panels in Fig.\ref{fig:imagenes} and right panel of Fig.\ref{fig:mono-cont}, respectively).
Along with the WR features and the broad emission line component, this confirms the presence of a circumnucleur starbust in CIG 993. 
The profiles near the  center of the galaxy are two to eight times brighter than the profiles in the rest of the galaxy.
The profiles on the northern side of the galaxy  have more counts than the profiles on the southern side, approximately 5 to 10\%.
This can also be seen in the number of counts in the right panel of Fig. \ref{fig:mono-cont}. The profile width also changes: the profile width in the northern side of the galaxy is larger 
than that in the southern half 
The monochromatic H$\alpha$ image shows a region of high H$\alpha$ intensity in the northern part of the galaxy, besides the already mentioned central parts of the galaxy.
This is probably a star formation region that is also visible in UV GALEX bands and optical images (Fig.\ref{fig:imagenes}).
In the velocity and residual map there is a peculiar  distribution of velocities along the major axis in the north side of the galaxy that could be interpreted as non-circular velocities associated with strong stellar winds in the star-forming region (marked E in Fig.\ref{fig:VIT-sigma} and \ref{fig:cr_Halpha}). 
Unfortunately the S/N on these parts of the galaxy does not allow us to do a further decomposition of the H$\alpha$ emission profiles.
CIG 993 shows several morphological and kinematical peculiarities.
The apparent disk of the galaxy looks  warped in optical (SDSS) and infrared (2MASS) images, but the residual velocity map does not show significant differences in velocity  over the southwestern side of CIG\,993 (values are actually close to zero).
Kinematically, there is no indication of warping over the whole disk.
Figure \ref{fig:disk-features} shows the DSS B-band of the galaxy with isophotes corresponding to the faintest emission.
The isophotes seem to trace an extended structure in the SW part of the disk, this feature looks like the beginning of an arm. 
A similar feature is seen from the isophotes on the composite SDSS image in Fig. \ref{fig:disk-features}.
This extended feature to the SW of the galaxy is probably not seen in SDSS optical or NIR bands because of the size of the telescopes and the respective exposure times.

\begin{figure}
\epsscale{0.8}
\plotone{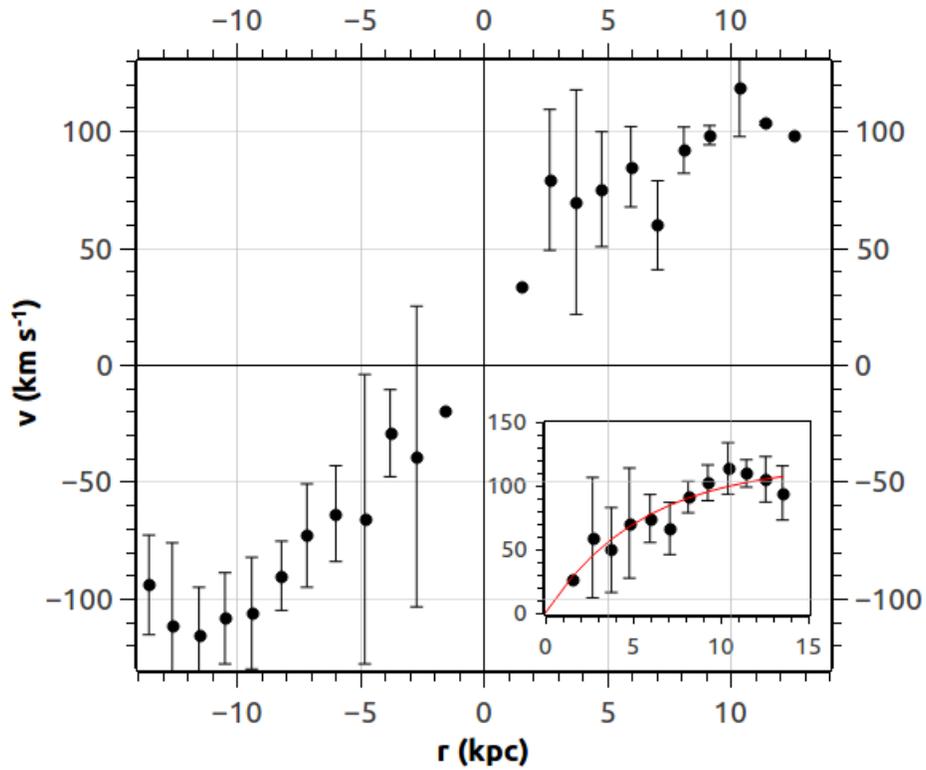} 
\caption{ Rotation curve for CIG 993 using the parameters of the ``H$\alpha$ RC'' in Section \ref{sec:RC} exept for a more face-on inclination $i=27^\circ$.
  \label{fig:RC-i27}}
\end{figure}

In order to explain the morpho-kinematics of CIG 993, we propose the following picture:
we believe CIG 993 is similar to the SB(s)c Seyfert 2 galaxy NGC 7479 (UGC 12343).
This galaxy has a very strong bar that dominates the surface profile with respect to the disk \citep{{Sanchez-Prtal2004}}. 
NGC 7479 also shows strong HII regions at one end of the bar \citep{Hernandez2005}.
The kinematics derived from FP scanning interferometry by \cite{Garrido2005} show that the velocities associated with the inner parts of NGC 7479 (within the inner $\sim 35 \arcsec$) slightly trace the rotation of the disk and that the amplitude associated to the RC in those inner parts corresponds to a fraction of the total V$_{max}$  of the galaxy -see their Figure A26.
 In the case of CIG 993, this would explain why the extension of the galaxy and the low rotation velocities do not seem to match; for a galaxy of such size and luminosity, higher velocities are expected.
 For CIG 993, considering the outermost isophotes of the DSS B-band and SDSS images leads to inclination values between $30^\circ$ and  $41^\circ$.
  These inclination values are considerably lower than those found in Section \ref{sec:RC} which result in $V_{max}$ values from the RC between 71 km s$^{-1}$ and 80 km s$^{-1}$.
 Another way of inferring the inclination value for a galaxy is through the Tully-Fisher relation (TFR). Using the luminosity (i.e. magnitude) data and the HI profile width, one is able to derive the inclination of a disk galaxy.
  We used the TFR derived from 2MASS bands following the methodology of \cite{Masters2008} which gives inclination values between $37^\circ$ and $41^\circ$ for J and K bands, respectively.
  Using the TFR for the $3.4\, \mu m$ WISE bands with the methodology of \cite{{Lagattuta2013}}, we find an inclination value of $i=27^\circ$.
  These results argue for a more face on configuration than the one inferred from shallower images of the galaxy in other wavelengths, as a consequence, a different RC can be derived where the maximum rotation velocity varies from 85 km s$^{-1}$ (considering $i=41^\circ$) to 126 km s$^{-1}$ (considering $i=27^\circ$). Fig. \ref{fig:RC-i27} shows the RC taking into account the same values of the ``H$\alpha$ RC'' except for the inclination value ($i=27^\circ$). We do not find evidence of a different position angle.
  The extension of the RC seems quite small (less than 14 kpc) respect to the H$\alpha$ and K-band RC and flatter than the other two.

 Using this value for the inclination of the disk of CIG 993, the \cite{Fernandez2010} and \cite{MirabelSanders1988} results on the HI line width (220.5 km s$^{-1}$ at 50\% and 215 km s$^{-1}$ at 20\%), and the \cite{Bottinelli1980} formula that  correlates them, we estimated a  $V_{max}$ value of 183 km s$^{-1}$.
  This value is 1.5 times larger than the one found with the optical RC.
  The difference between the $V_{max}$ values might be related to the fact that the TFR makes use of the HI information which is usually more extended than the information from the ionized gas. In particular for this galaxy, the ionized gas is confined to the central parts of the galaxy and the inner parts of the disk, thus the optical RC is tracing less mass than the HI RC. Also the flat part of the optical RC is not reached.
 Since neutral hydrogen emission can be detected over almost the entire radial extent of a spiral galaxy, 21-cm line data is a more powerful source of information on the kinematics and thus the dynamics of the galaxy as a whole, while the interferometric observations of the ionized gas give important information on the different kinematical components in the inner parts of CIG 993 where the outflows are being detected.



%
What we are probably observing in this case is the central parts of a galaxy similar to NGC 7479 but five times further away, so that only the bulge, the bar and the beginning of the disk -including one of the spiral arms- are seen.
Contrary to CIG 993, NGC 7479 does not harbor a circumnuclear starburst, but is classified as a Sy2 galaxy.
This could indicate that CIG 993 is in a previous stage in the evolution from a LIRG galaxy to an active galaxy as shown by several authors \citep[e.g.][]{Sanders1988,Hopkins2008}.
CIG 993  is  probably a late type spiral because of the high star formation rate, like the LCBG sample of \citet{MallenOrnelas1999}, rather than Sa (RC3).

According to \cite{Armus1988} the high number of WR stars in relation to the number of O-type stars (inferred from the flux ratio of the $4686 {\rm\AA}$ and $H\beta$ features) may argue against a continuous  steady state star formation in CIG\,993, and in favor of recent burst of star formation occurring approximately $10^7$ yr ago. 
 On the other hand  if the galaxy is isolated (according to the Karachentseva criterion) then it has  a lower time limit for a possible past interaction, that is around $10^9$ yr.
 The fact that CIG 993 is classified as an e(c) galaxy imposes a lower time limit for the starburst duration of  $10^8$ yr, as long as the SF continues, which is the case for this galaxy.
The two last arguments point to a long star-forming process if induced by an interaction. 
 Finally, the estimation of the number of WR stars in the galaxy might have been over-estimated because of an ``artificial'' widening of the emission profiles due to the central outflow in the galaxy.
In such a case, a possible past interaction could have triggered the star formation in CIG\,993 and originate the LIRG, LCBG, WR galaxy, dusty starburst, e(c) features, the asymmetrical HI profile \citep{Fernandez2010} and the central outflow.
Also, outflows have serious impact on the structure and chemical composition of the interstellar medium and mass distribution being also  the primary mechanism by which dust and metals are distributed over large scales within the galaxy transferring the kinetic energy and starting the processes of star formation \citep{Veilleux2005},  as a consequence they contribute to the metal enrichment of the intergalactic medium. This scenario could explain the metallicity of $12+log(O/H)\sim 9.3-9.4$ in CIG 993  \citep{Schaerer2000}.

\section{Conclusion} 
We present the kinematic analysis of the  ionized gas of the galaxy CIG\,993, using  the scanning Fabry-Perot interferometer PUMA.
Analysis of the velocity map reveals a gradient from northwest to southeast, indicating  the presence of rotational motions in the galaxy.
However important departures from circular motion in the central region of the galaxy  are detected on the VF, the $\sigma$ map and the residual velocity map of the galaxy.
The spectral resolution of PUMA allowed us to decompose the H$\alpha$ emission profiles in this region  and identify both a narrow and a broad components.
There is a strong concentration of H$\alpha $ emission in the central zone of the galaxy where strong massive star formation is taking place, marked by high number of  WR and O stars.

The broad band component could thus be associated with the presence of an outflow most likely related to the LBCG and e(c) nature of the galaxy, as well as the presence of WR features.
%
In the north side of the galaxy there is a zone of high star formation where both the VF and residual velocities map indicate the presence of non-circular motions probably associated with stellar winds from young massive stars. 
The morphology of CIG 993 is  difficult to define. The direct images in different bands indicate the presence of a bulge, a disk-like feature seen with a high inclination and a warp.
However kinematically it is difficult to associate V$_{max}$ from the derived RC with the apparent size of the galaxy. 
Plus, the DSS B-band and SDSS composite color images show an interesting feature in the southwestern side of the disk (see Fig.\ref{fig:imagenes}), where instead of a warp, the emission seems to trace the beginning of a spiral arm. 
We think that CIG\,993 is most likely a galaxy similar to NGC 7479, an SBc galaxy with a very bright bar and HII regions both in the inner and outer parts of the galaxy. Since CIG 993 is located five times further away than NGC 7479, neither H$\alpha$ emission or emission in other bands is detected from the underlying disk. The fact that we are only detecting ionized gas from the central parts of the galaxy results in a low velocity RC  for different inclinations, centers and P.A. The present analysis shows that the structure of CIG993 is different when considering the fainter outer regions, revealing a more face-on galaxy. Authors as \cite{Davies1973} have already mentioned the important of the outer regions in the determination of the axis ratio as well as the inclination of disk galaxies.

%
Although CIG\,993 is an isolated galaxy, the AMIGA project  found  that  about  92\%  of  neighbor galaxies that show recession velocities similar to the corresponding CIG galaxy are not considered in the CIG neighbors  catalog \citep{Verley2007} because these neighbors are most likely dwarf systems, with less than a quarter diameter of the AMIGA galaxies'. However, interactions with this type of galaxies may have had considerable influence on the evolution of CIG galaxies \citep{Argurdo-Fernandez2013}.
Additionally, the asymmetric HI profile  reported by \cite{MirabelSanders1988} and \cite{Fernandez2010}  might be linked to recent accretion events  taking place  between $10^7$ to $10^9$ yr ago \citep{Osterbrock&Cohen1982,AMIGA05}. 
 This results suggest a  gravitational perturbation  could have triggered the star formation  responsible for the central outflow in CIG\,993, as well as the LIRG, LCBG, WR galaxy, dusty starburst and e(c) features.
 If this is the case, this interaction -probably a small satellite accretion- did not leave either a morphological or kinematical imprint, but could still ``ignite'' long-lived SF events given the galaxy had enough gas supply and the correct encounter parameters.

\acknowledgments

 I.F-C. thanks the financial support from the IPN-SAPPI project  20181136. N.C-M. acknowledges CONACyT for a doctoral scholarship.

\vspace{5mm}





\begin{thebibliography}{}





\bibitem[Afanas'ev et al.(1980)]{Afanasev1980} Afanas'ev, V.~L., Lipovetskii, V.~A., Markaryan, B.~E., \& Stepanyan, D.~A.\ 1980, Astrophysics, 16, 119 

\bibitem[Alonso-Herrero et al.(2012)]{Alonso-Herrero2012} Alonso-Herrero, A., Pereira-Santaella, M., Rieke, G.~H., \& Rigopoulou, D.\ 2012, \apj, 744, 2 


\bibitem[Arribas et al.(2014)]{Arribas2014} Arribas, S., Colina, L., Bellocchi, E., Maiolino, R., \& Villar-Mart{\'{\i}}n, M.\ 2014, \aap, 568, A14 

\bibitem[Argudo-Fern{\'a}ndez et al.(2013)]{Argurdo-Fernandez2013} Argudo-Fern{\'a}ndez, M., Verley, S., Bergond, G., et al.\ 2013, \aap, 560, A9 

\bibitem[Armus et al.(1988)]{Armus1988} Armus, L., Heckman, T.~M., \& Miley, G.~K.\ 1988, \apjl, 326, L45


\bibitem[Begeman(1989)]{Begeman1989} Begeman, K.~G.\ 1989, \aap, 223, 4


\bibitem[Bottinelli et al.(1980)]{Bottinelli1980} Bottinelli, L., Gouguenheim, L., Paturel, G., \& de Vaucouleurs, G.\ 1980, \apjl, 242, L153 

\bibitem[Brassington et al.(2015)]{Brassington2015} Brassington, N.~J., Zezas, A., Ashby, M.~L.~N., et al.\ 2015, \apjs, 218, 6 


\bibitem[Davies(1973)]{Davies1973} Davies, R.~D.\ 1973, \mnras, 161, 25P 


\bibitem[de Grijp et al.(1985)]{deGrijp1985} de Grijp, M.~H.~K., Miley, G.~K., Lub, J., \& de Jong, T.\ 1985, \nat, 314, 240 

\bibitem[de Vaucouleurs et al.(1991)]{rc31991} de Vaucouleurs, G., de Vaucouleurs, A., Corwin, H.~G., Jr., et al.\ 1991, Third Reference Catalogue of Bright Galaxies.

\bibitem[Dennefeld et al.(1986)]{Dennefeld1986} Dennefeld, M., Karoji, H., \& Belfort, P.\ 1986, Star-forming Dwarf Galaxies and Related Objects, 351 


\bibitem[Dressler et al.(1999)]{Dressler1999} Dressler, A., Smail, I., Poggianti, B.~M., et al.\ 1999, \apjs, 122, 51 


\bibitem[Espada et al.(2011)]{Espada2011} Espada, D., Verdes-Montenegro, L., Huchtmeier, W.~K., et al.\ 2011, \aap, 532, A117 















\bibitem[Fernandes et al.(2004)]{Fernandes2004} Fernandes, I.~F., de Carvalho, R., Contini, T., \& Gal, R.~R.\ 2004, \mnras, 355, 728 


\bibitem[Fernandez et al.(2010)]{Fernandez2010} Fernandez, M.~X., Momjian, E., Salter, C.~J., \& Ghosh, T.\ 2010, \aj, 139, 2066 

\bibitem[Fuentes-Carrera et al.(2004)]{Iqui2004} Fuentes-Carrera, I., Rosado, M., Amram, P., et al.\ 2004, \aap, 415, 451 

\bibitem[Garrido et al.(2005)]{Garrido2005} Garrido, O., Marcelin, M., Amram, P., et al.\ 2005, \mnras, 362, 127 


\bibitem[Garland et al.(2004)]{Garland2004} Garland, C.~A., Pisano, D.~J., Williams, J.~P., Guzm{\'a}n, R., \& Castander, F.~J.\ 2004, \apj, 615, 689

\bibitem[Genzel et al.(1998)]{Genzel1998} Genzel, R., Lutz, D., Sturm, E., et al.\ 1998, \apj, 498, 579 



\bibitem[Haan et al.(2011)]{Hann2011} Haan, S., Surace, J.~A., Armus, L., et al.\ 2011, \aj, 141, 100 



\bibitem[Helou et al.(1985)]{Helou1985} Helou, G., Soifer, B.~T., \& Rowan-Robinson, M.\ 1985, \apjl, 298, L7 


\bibitem[Hernandez et al.(2005)]{Hernandez2005} Hernandez, O., Carignan, C., Amram, P., Chemin, L., \& Daigle, O.\ 2005, \mnras, 360, 1201 


\bibitem[Hopkins et al.(2008)]{Hopkins2008} Hopkins, P.~F., Hernquist, L., Cox, T.~J., Dutta, S.~N., \& Rothberg, B.\ 2008, \apj, 679, 156 




\bibitem[Ishida (2004)]{Ishida2004}  Ishida, C. M.  2004, Thesis (PhD) 



\bibitem[Karachentseva (1973)]{Kara73}Karachentseva, V. E. 1973, Soobshch. Spets. Astrofiz. Obs, 8, 72


\bibitem[Kelz et al.(2006)]{Kelz2006} Kelz, A., Verheijen, M.~A.~W., Roth, M.~M., et al.\ 2006, \pasp, 118, 129 


\bibitem[Kennicutt et al.(1987)]{Kennicutt1987} Kennicutt, R.~C., Jr., Roettiger, K.~A., Keel, W.~C., van der Hulst, J.~M., \& Hummel, E.\ 1987, \aj, 93, 1011 



\bibitem[Kennicutt(1998)]{Kennicutt98} Kennicutt, R. C., Jr. 1998, \araa, 36, 189


\bibitem[Kunth \& Schild(1986)]{Kunth&1986} Kunth, D., \& Schild, H.\ 1986, \aap, 169, 71 


  \bibitem[Lagattuta et al.(2013)]{Lagattuta2013} Lagattuta, D.~J., Mould, J.~R., Staveley-Smith, L., et al.\ 2013, \apj, 771, 88 


 \bibitem[Lequeux(1983)]{Lequeux1983} Lequeux, J.\ 1983, \aap, 125, 394 
 
 
\bibitem[Leitherer et al.(2013)]{Leitherer2013} Leitherer, C., Chandar, R., Tremonti, C.~A., Wofford, A., \& Schaerer, D.\ 2013, \apj, 772, 120


\bibitem[Lisenfeld et al.(2007)]{Lisenfeld2007} Lisenfeld, U., Verdes-Montenegro, L., Sulentic, J., et al.\ 2007, \aap, 462, 507 




\bibitem[Mall{\'e}n-Ornelas et al.(1999)]{MallenOrnelas1999} Mall{\'e}n-Ornelas, G., Lilly, S.~J., Crampton, D., \& Schade, D.\ 1999, \apjl, 518, L83 

 
  \bibitem[Masters et al.(2008)]{Masters2008} Masters, K.~L., Springob, C.~M., \& Huchra, J.~P.\ 2008, \aj, 135, 1738 


\bibitem[Mirabel \& Sanders(1988)]{MirabelSanders1988} Mirabel, I.~F., \& Sanders, D.~B.\ 1988, \apj, 335, 104 

\bibitem[Melbourne et al.(2008)]{Melbourne2008} Melbourne, J., Ammons, M., Wright, S.~A., et al.\ 2008, \aj, 135, 1207 

\bibitem[Melbourne et al.(2005)]{Melbourne2005} Melbourne, J., Koo, D.~C., \& Le Floc'h, E.\ 2005, \apjl, 632, L65

 \bibitem[Nordsieck(1973)]{Nordsieck1973} Nordsieck, K.~H.\ 1973, \apj, 184, 719 

\bibitem[O'Halloran et al.(2005)]{OHalloran2005} O'Halloran, B., McBreen, B., Metcalfe, L., Delaney, M., \& Coia, D.\ 2005, \aap, 439, 539 


\bibitem[Oparin \& Moiseev(2015)]{OparinMoiseev2015} Oparin, D.~V., \& Moiseev, A.~V.\ 2015, Astrophysical Bulletin, 70, 392 

\bibitem[Osterbrock(1989)]{Osterbrock1989} Osterbrock, D.~E.\ 1989, Research supported by the University of California, John Simon Guggenheim Memorial Foundation, University of Minnesota, et al.~Mill Valley, CA, University Science Books, 1989, 422 p., 

\bibitem[Osterbrock \& Cohen(1982)]{Osterbrock&Cohen1982} Osterbrock, D.~E., \& Cohen, R.~D.\ 1982, \apj, 261, 64 

\bibitem[Osterbrock \& Dahari(1983)]{OsterbrockDahari1983} Osterbrock, D.~E., \& Dahari, O.\ 1983, \apj, 273, 478 

 \bibitem[Osterbrock et al.(1996)]{Osterbrock1996} Osterbrock, D.~E., Fulbright, J.~P., Martel, A.~R., et al.\ 1996, \pasp, 108, 277 
 



\bibitem[Pisano et al.(2001)]{Pisano2001} Pisano, D.~J., Kobulnicky, H.~A., Guzm{\'a}n, R., Gallego, J., \& Bershady, M.~A.\ 2001, \aj, 122, 1194  

  

\bibitem[P{\'e}rez-Gallego et al.(2011)]{Perez-Gallego2011} P{\'e}rez-Gallego, J., Guzm{\'a}n, R., Castillo-Morales, A., et al.\ 2011, \mnras, 418, 2350 

\bibitem[Poggianti et al.(1999)]{Poggianti1999} Poggianti, B.~M., Smail, I., Dressler, A., et al.\ 1999, \apj, 518, 576 

\bibitem[Poggianti \& Wu (2000)]{PoggiantiWu2000} Poggianti, B.M. \& Wu, H. 2000, \apj , 539, 157  


\bibitem[Rosado et al. (1995)]{Rosado1995} Rosado, M., Langarica, R., Bernal, A., Cobos, F., et al. 1995, Revista Mexicana de Astronomia y Astrofisica Conference Series, 3, 264 


\bibitem[Roth et al.(2005)]{Roth2005} Roth, M.~M., Kelz, A., Fechner, T., et al.\ 2005, \pasp, 117, 620 


\bibitem[Sabater et al.(2008)]{Sabater2008} Sabater, J., Leon, S., Verdes-Montenegro, L., et al.\ 2008, \aap, 486, 73

 
\bibitem[Sanders \&  Mirabel(1996)]{S&M1996} Sanders, D. B. \&  Mirabel, I. F.  1996, \araa,  34, 746


\bibitem[Sanders et al.(1988)]{Sanders1988} Sanders, D.~B., Soifer, B.~T., Elias, J.~H., et al.\ 1988, \apj, 325, 74 

 
\bibitem[S{\'a}nchez-Portal et al.(2004)]{Sanchez-Prtal2004} S{\'a}nchez-Portal, M., D{\'{\i}}az, {\'A}.~I., Terlevich, E., \& Terlevich, R.\ 2004, \mnras, 350, 1087 

\bibitem[Schaerer et al.(2000)]{Schaerer2000} Schaerer, D., Guseva, N.~G., Izotov, Y.~I., \& Thuan, T.~X.\ 2000, \aap, 362, 53 

\bibitem[Spitzer(1978)]{Spitzer1978} Spitzer, L.\ 1978, Physical processes in the interstellar medium, by Lyman Spitzer.~ New York Wiley-Interscience, 1978.~333 p., 



\bibitem[Sulentic et al.(2006)]{Sulentic2006} Sulentic, J.~W., Dultzin-Hacyan, D., Marziani, P., et al.\ 2006, \rmxaa, 42, 23 









\bibitem[Veilleux et al.(2005)]{Veilleux2005} Veilleux, S., Cecil, G., \& Bland-Hawthorn, J.\ 2005, \araa, 43, 769 


\bibitem[Veilleux et al.(1995)]{Veilleux1995} Veilleux, S., Kim, D.-C., Sanders, D.~B., Mazzarella, J.~M., \& Soifer, B.~T.\ 1995, \apjs, 98, 171 




\bibitem[Verdes-Montenegro et al.(2005)]{AMIGA05} Verdes-Montenegro, L., Sulentic, J., Lisenfeld, U., Leon, S., Espada, D., Garcia, E., Sabater, J.  \& Verley, S.  2005, \aap,   436, 443

 \bibitem[Verley et al.(2007)]{Verley2007} Verley, S., Odewahn, S.~C., Verdes-Montenegro, L., et al.\ 2007, \aap, 470, 505




\bibitem[Warner et al.(1973)]{Warner1973} Warner, P.~J., Wright, M.~C.~H., \& Baldwin, J.~E.\ 1973, \mnras, 163, 163 

\bibitem[Wu et al. (1998)]{Wu1998} Wu, H., Zou, Z. L., Xia, X. Y. \& Deng, Z. G. 1998, A\&AS , 132, 181


\bibitem[Yuan et al.(2010)]{Yuan2010} Yuan, T.-T., Kewley, L.~J., \& Sanders, D.~B.\ 2010, \apj, 709, 884 





\end{thebibliography}
\end{document}